\documentstyle{article}
\begin{document}
\begin{center}
{Nuovo Cimento D 20 (1998) 73-78 \& cond-mat/9608014}
\end{center}
\begin{center}{\Large{\bf Why DNA ?}\\

H. Rosu
\footnote{e-mail: rosu@ifug.ugto.mx

Address before July 1990: IGSS, Magurele-Bucharest, Romania}\\[2mm]
{\scriptsize{Instituto de F\'{\i}sica, Universidad de Guanajuato,\\
Apdo Postal E-143, 37150 Le\'on, Gto, M\'exico}\\  } }
\end{center}


\begin{center}
{\bf Abstract}

A small collection of general facts related to DNA is presented.

\vskip 0.5cm
\end{center}





\begin{center} {\bf Introduction} \end{center}

The DNA genetic macromolecule is certainly a remarkable entity of
the Universe. First of all, it is a compact carrier of an astounding
amount of biological information. One fundamental goal concerning DNA might
be to
reveal general mathematical aspects that may be related to its structure
and thereby finding
reasons for it to exist. Such general facts can be found in different
mathematical disciplines and they must be carefully collected with the
hope of providing further interesting insights on DNA macromolecules.
For example, interesting topological considerations as applied to DNA
are quite well-known \cite{1}. At the present time there is much
publishing activity in two areas - (i) nonlinear dynamical models of DNA,
for a review see \cite{g},
and (ii) statistical
correlations of bases and noncoding regions as well as defining measures of
the informational content of the genetic molecule \cite{ep}.

The sections of this paper are as follows.
First, the most general features of DNA structure are presented
for self-consistency reasons.
Second, several comments on the partial differential equations approach
are introduced.
Thirdly, I quote from two remarkable books on possible differential
geometric arguments for
the helix structure of DNA. They apply to the first level of the hierarchy
of DNA models in Yakushevich classification \cite{y94}, i.e., to the
{\em thin elastic filament} paradigm and its discrete version.
This is the simplest model that nevertheless
may be of relevance to the folding problem. Indeed, supposing the genetic
acids be completely stable chemically, one may look at
them not as heteropolymers, but as elastic thin rods. This is especially
appropriate
for investigating the basic features of their motions, such as their
foldings inside cells. DNA foldings were not
considered from this standpoint, and precisely in this regard
I wish to point out the known, remarkable
connections between one-dimensional soliton equations and the geometrical
motion of space curves \cite{mat}. In other words, the conjecture is that
solitons of chemical origin living on the moving DNA curves are
responsible for the folding.

\begin{center} {\bf The DNA macromolecule} \end{center}

The famous double-helical structure of the DNA (Watson and Crick \cite{cw})
shows that a great deal of the essence of life belongs to the remarkable
{\em helicoidal world} \cite{d93}.

DNA is essentially made of two polymer backbones of interwound
strands \cite{f}. The subunits of the polymer backbones are known as
nucleotides. In each backbone, phosphate groups alternate with the
deoxyribose sugar, resulting in a covalently linked chain. The chains have
polarity or direction, i.e., in the duplex they run
parallel but in opposite directions, with a right-handed twist.

To the sugar ring of each nucleotide one will find attached in an
apparently random way one of the following four bases: adenine, guanine,
thymine and cytosine (symbolically A,G,T,C). In the case of RNA,
thymine is substituted by uracil (U).
The first two bases are purines, whereas the last two are pyrimidines.
The information about the composition of all the protein molecules
in one organism is given by the order of the nucleotide bases along
the DNA chain. Each base is linked by either {\em two or three}
hydrogen bonds to a base on the opposite strand. The pairing of the strands
is almost exclusively A$=$T (A$=$U for RNA) and {\em G$\equiv$C}, which is the
Watson-Crick base pairing. This is a consequence of the
chemical Chargaff's rules: in nucleic acids,
the molar quantities of A and T are equal, as well as of G and C.
It is only the molar ratio purines/pyrimidines that is varying among
species.

The point I would like
to emphasize here is that both the {\em sugar configurations} as well as the
{\em bases} are {\em pentagonal configurations}, for which the ratio of
the diagonal to the side is exactly the golden mean. Therefore,
the golden mean is strongly present in the DNA architecture, in a direct,
exact way and not as a Fibonacci limit that I am sure be there as well,
probably in some base correlations.

\begin{center}
{\bf Partial differential equations: what could they tell us about
DNA ?}
\end{center}

The main mathematical technique of studying the dynamics of DNA is
through {\em classical} partial differential equations, because they are
governing the properties of the mesoscopic and macroscopic
order parameters and thus they may be considered as approximately
describing various nucleation processes in a dynamic way.
Since DNA is a very complicated
molecule, one expects the application of nonlinear differential equations
to be overwhelming. And indeed in the last ten years or so, one has
witnessed many studies regarding DNA solitons, mainly as extensions of
works on molecular chains. Yomosa, following a suggestion of
Englander {\em et al} \cite{en80}, was one of the first people
to introduce firmly sine-Gordon solitons for DNA double helices \cite{y83},
and he was able to prove that major dynamical processes, like replication,
transcription, and mutation are well described by means of various
couplings between kinks. Actually, Englander {\em et al}
proved by experiment the existence of so-called mobile open units of the
order of ten base pairs in length diffusing along the double helix.
For a recent review on DNA solitonic motion, see Gaeta {\em et al} \cite{g}.
One may ask why soliton excitations might be of relevance for DNA. A partial
answer is as follows.
Among the most general non-linear partial differential equations (PDE)
those of reaction-diffusion type are used to model many types of
spatiotemporal patterns and their dynamics.
Especially important in this class of PDE are the time-dependent
Landau-Ginzburg equations
\begin{equation}\label{eq:dna2}
\frac{\partial \eta}{\partial t}=D\nabla ^{2}\eta +P(\eta)~.
\end{equation}
D is a diffusion coefficient, $P(\eta)$ is a polynomial whose
particular form depends on the details of the model and the quantity
$\eta$ is what one usually calls an order parameter helping to
distinguish two opposing (or of different symmetry) phenomena that can be
considered of dynamical or thermodynamical origin. Also, noises can be included
in various ways. Some time ago, Skierski,
Grundland, and Tuszy\'nski \cite{sgt} provided a systematic study
showing how to bring a PDE of the form given in Eq.(\ref{eq:dna2})
to the form of an anharmonic-oscillator equation (AOE), i.e.,
$\ddot u +\gamma \dot u = P(u)$, by means of
symmetry reduction procedures. On the other hand, AOE has
well-established interpretations and are very well known
mathematically. It appears that in the case of Ginzburg-Landau PDE
there are two types of symmetry reductions. The first is when the symmetry
variable is $\xi =x-vt$, {\em i.e.}, travelling-wave solutions.
Solitons and kinks belong to this class of straight line propagation
of constant velocity.
The second class is given by the polar choice $\xi =\phi -\log r^{b}$,
where $\phi$ is the azimuthal angle, $r$ is the radial variable
in the $(x,y)$ plane and $b$ is a scaling parameter. This corresponds
to spiral patterns of the order parameter and shows why such patterns are
so common in excitable media \cite{spir}.

\begin{center} {\bf Homogeneous curves and their self-motions}
\end{center}

Keeping in mind the above classes of curves corresponding to the order
parameter under GL symmetry reductions, I now recall some general
geometric properties of homogeneous curves just by quoting (with a few addings
only) from two famous books \cite{ya} and \cite{eu}.


(I).- Homogeneous curves are those having {\em all their points the same}.
In plane Euclidean geometry there are only two types of homogeneous
curves, the {\em straight lines} ($\cal L$) and the {\em circles}
($\cal C$).
Such curves have
self-isometry groups of motion, i.e., sets of isometries sending the
 curve into itself, or in other words, each of its points into another
of its points. In the case of straight lines, the self-motions (the
group of motions along itself) is the group of translations in the
direction of the line. In the case of the circle, the self-motions
are the group of rotations about the center of the circle.

There exists only one more plane curve which has almost the same
complete
degree of homogeneity as the straight line and the circle. This is
the {\em logarithmic spiral} ($\cal L$$\cal S$). In polar coordinates
$r,\;\phi$, the {$\cal L$$\cal S$ equation
is $r=a^{\phi}$, where $a$ is related to the initial radius and the
coefficient of growth (or collapse) of the spiral.
It is well-known that $\cal L$$\cal S$ has similarity
motions along itself,
sending $\cal L$$\cal S$ into itself, and therefore every point on it
into another point onto it.
The similarity motions of $\cal L$$\cal S$ in polar coordinates are
$r^{'}=a^{c}r$, and $\phi ^{'}=\phi +c$, where $c$ is a constant parameter
(angle).


(II).-
Old Greek Geometers payed sufficient attention to a special type of curves
known under the name of {\em homoeomeric} lines, i.e.,
{\em lines which are alike in all parts, so
that in any one such curve any part can be made to coincide with any
other part}. Proclus remarked that these lines are only three, two of
which are simple and in a plane, and the third mixed, namely the cylindrical
helix (called also {\em cochlias} or {\em cochlion}).
Its homoeomeric property was
proven by Apollonius, and the fact that there are only three classes of
homoeomeric lines was proved by Geminus, who also proved the following
general property: if from a point two straight lines are to be drawn to a
homoeomeric line making equal angles with it, the straight lines are equal.
Proclus considered the cochlion as a mixed line, in the sense
that it is not generated by two simple/like motions (i.e., of the same
{\em speed}).
Geminus explained that the term ``mixed'' as applied to curves, and as applied
to surfaces, respectively, is used in different senses. For curves, mixing is
neither ``putting together" nor ``blending". If cut in any way, it does not
present the appearance of any of the simple lines of which it is made of.

\begin{center} {\bf Conclusion} \end{center}

The double helix structure of DNA is, roughly speaking, the combined result
of chemistry and the tendency to the {\em cochlion} configuration
(two or more interwound helices fill more completely a cylindrical space).

The interplay between the nucleation processes, the general, `geometric'
principles dictating the configurations of important macromolecules and
the kinematic potentiality as a function of the environment is
clearly an open issue, since most people get used to think only in
energetic (variational) way.

A powerful mathematical argument in favor of the helical structure of
the DNA is the self-similar property of the helical curves. However,
DNA is not a perfect self-similar organization and one should introduce
methods to describe this approximate self-similarity \cite{jap}.

Let us mention also that the classical analog of
Schr\"odinger's Zitterbewegung is the helical motion.
There is a revival of interest in studying
this analogy. Recently, Pav\v si\v c {\em et al} \cite{Pa92} showed
that the electron is associated to the mean motion of point-like constituents,
whose trajectory is, at the classical level, a cylindrical helix.
The occurence of Zitterbewegung can be explained by the general
self-similar properties of the helical motions, making them to come into
being as {\em kinematical germs} at any spatial scale.
For example, at the level of the electroweak theory,
Poppitz \cite{popp} showed that a finite fermion density
destabilizes the W-pair condensate in a strong, external magnetic field, and
breaks the translational invariance in the direction of the field, leading to
the appearance of a helicoidal structure.

\begin{center} *** \end{center}

This work was partially supported by the CONACyT Project 4868-E9406.


\end{document}